# Thermal diffusion of ionic species in charged nanochannels

Wei Qiang Chen*, Majid Sedighi*, and Andrey P Jivkov

Department of Mechanical, Aerospace and Civil Engineering, School of Engineering, The University of Manchester, Manchester, M13 9PL, United Kingdom

*Corresponding authors: Weiqiang.Chen@manchester.ac.uk & Majid.Sedighi@manchester.ac.uk

**Abstract**

Diffusion of ions due to temperature gradients (known as thermal diffusion) in charged nanochannels is of interest in several engineering fields, including energy recovery and environmental protection. This paper presents a fundamental investigation of the thermal diffusion of sodium chloride in charged silica nanochannels performed by molecular dynamics (MD). The results reveal the effects of nanoconfinement and surface charges on the sign and magnitude of the Soret coefficient. It is shown that the sign and magnitude of the Soret coefficient are controlled by the structural modifications of the interfacial solutions. These modifications include the ionic solvation and hydrogen bond structure induced by the nanoconfinement and surface charges. The results show that both nanoconfinement and surface charges can make the solutions more thermophilic. Furthermore, the thermal diffusion of solutions in boundary layers is significantly different from that of solutions in bulk fluid, contributing to the overall difference between the thermal diffusivity of pore fluid and that associated with bulk fluid. The findings provide further understanding of thermal diffusion in nano-porous systems. The proposed MD simulation methodology is applicable to a wider category of coupled heat and mass transfer problems in nanoscale spaces.

# 1. Introduction

Thermal diffusion in fluid mixtures and thermophoresis in suspensions, which are also referred to as the Ludwig-Soret effect, describe the migration of solutes due to temperature gradients. Thermal diffusion participates in various natural processes and is of interest for a range of engineering applications such as thermal field-flow fractionation of colloids and synthetic polymers [1, 2], spatial composition analysis of hydrocarbon reservoirs [3], combustion processes [4], and behaviour of solutions encompassing ions and polar solutes. It has been shown that the thermal diffusion of biomolecules is highly sensitive to their structural modifications in aqueous solutions. This has been used to monitor the protein-ligand binding reactions under temperature gradients [5]. Thermal diffusion can also be used to microscopically manipulate the motions of biomolecules [6] and colloidal particles, which can work as nanocarrier agents for drug delivery [7]. In low carbon energy conversion, understanding the thermo-diffusive response of electrolyte solutions and other charged fluids is particularly useful in quantifying the thermoelectric Seebeck effect. This is being extensively studied to develop devices capable of recovering waste heat or solar energy and converting it to electricity [8-12]. Further, evaluating the thermal diffusion of geofluids is vital for the long-term safety of many geotechnical infrastructures such as nuclear waste repositories, landfill containers, and geothermal energy facilities [13, 14].

Despite significant developments in the understanding of thermal diffusion and the successful utilization of this physical phenomenon in aqueous systems in many fields, a full microscopic explanation of the phenomenon is missing. As a result, there is no general theory that can successfully evaluate the magnitude and direction of diffusive flux due to thermal diffusion in aqueous solutions in a wide range of situations. A number of experimental observations have revealed peculiar phenomena, such as a change of the sign of the Soret coefficient at a specific temperature (a negative coefficient means that migration occurs from colder to hotter regions) and the existence of a minimum Soret coefficient [15]. Application of different membranes, microfluidic and nanofluidic devices by utilizing thermal diffusion of aqueous solutions, e.g., thermophoretic molecule trap for DNA replication and accumulation [16], and temperature-controlled nano-porous membrane for molecular transport and characterization [17] have emerged. Such technological developments highlight the need for further understanding of thermal diffusion in microscale and nanoscale spaces and channels. It has been suggested in the literature that in addition to the nano-confinement effect on thermal diffusion, the effect of charged solid surfaces should also be considered, as it exists in various industrial applications and natural systems, e.g., negatively/positively charged nano-porous membranes for ion transport and ionic selectivity [18], osmotic energy harvesting based on pressure-retarded osmosis and reversed electrodialysis methods [19], electric energy storage technologies such as alkaline zinc-based aqueous flow batteries [20], clay minerals forming the nanochannels for geofluids are mostly negatively charged [21]. While the understanding of the thermo-diffusive response of bulk aqueous solutions has been improved recently by experimental, theoretical, and numerical studies, e.g., [6, 13, 14, 22-24], the research on thermal diffusion in nano-confined aqueous solutions has been very limited, particularly in charged media. It is known that nanoconfinement and surface charges will significantly change the transport properties of the interfacial liquid, e.g., diffusive transport [25-29]. Therefore, the existing knowledge on the thermo-diffusive properties of bulk liquid cannot be directly applied.

The aim of this work is to advance the mechanistic understanding of thermal diffusion in charged nanoscale structures. To the best of our knowledge, the combined effects of nanoconfinement and surface charges on thermal diffusion have not been studied from a molecular-level perspective before, whilst such understanding is critical to facilitate the technological and scientific advances in the areas mentioned above. To this end, the thermal diffusion of NaCl aqueous solutions confined in charged quartz nanochannels is investigated by equilibrium molecular dynamics (EMD) and non-equilibrium molecular dynamics (NEMD) simulations. The choice of the solid phase is dictated by practical considerations. Silicon-based nanomaterials are appropriate for device applications due to their superb biocompatibility, biodegradability, and unique optical, electronic, and mechanical properties [30]. These make quartz and other silica phases vital constituents of micro/nanodevices for water filtration [31-34], drug delivery [30], biomolecule detection [35, 36], etc. The paper is organized as follows. The theory and simulation approaches are discussed in Section 2. The results of the simulations are presented in Section 3. They show how the thermo-diffusive response of the solutions is affected by the solution-quartz interactions and the nanoconfinement. The findings are explained by the alteration of the interfacial liquid structure, which is also discussed in Section 3. Finally, conclusions are drawn in Section 4.

## 2. Theory and method

Thermal diffusion and the associated MD systems to measure the Soret effect of sodium chloride (NaCl) in water are presented in this section. Aqueous NaCl solutions are binary mixtures consisting of salt (component 1, solute) and water (component 2, solvent).

### 2.1 Thermodynamic formulation and molecular dynamics setup

Based on non-equilibrium thermodynamics [37], the diffusive mass flux of solute (component 1), $J_1$, in a non-reacting binary mixture under a thermal gradient and in mechanical equilibrium is given by [38]

$$J_1 = -\rho D_{12} \nabla w_1 - \rho w_1 (1 - w_1) D_T \nabla T, \qquad (1)$$

where $D_T$ and $D_{12}$ are the thermal and the effective mutual mass (Fickian) diffusion coefficients of the mixtures, respectively, $w_1$ is the weight fraction of component 1, $\rho$ is the mass density of the mixture, and $T$ is the temperature. Under a given temperature gradient, the system tends towards an equilibrium state of zero mass flux. The Soret coefficient, generally used to quantify thermal diffusion, characterises the equilibrium state. In the case of dilute solutions, it is evaluated by:

$$S_T = \frac{D_T}{D_{12}} = -\frac{1}{w_1(1-w_1)}\left(\frac{\nabla w_1}{\nabla T}\right)_{J_1=0} = -\frac{1}{x_1(1-x_1)}\left(\frac{\nabla x_1}{\nabla T}\right)_{J_1=0} \approx -\frac{1}{x_1}\left(\frac{dx_1}{dT}\right)_{J_1=0}, \qquad (2)$$

where $x_1$ is the molar fraction of component 1 (solute). This equation is an approximation, as it is derived with the assumption that the quantity of solvent significantly exceeds the quantity of the solute. Thus, a positive $S_T$ implies that the solute tends to move with respect to the solvent from the hotter to the colder region. Notably, the existing experimental approaches do not explicitly distinguish the Soret coefficient for the cations and anions and provide Soret coefficients of individual ions. In this work, it is assumed that $S_{T,Na^+} = S_{T,Cl^-} = S_T$, which is also assumed in previous MD simulations [15].

Previous studies have found that the following empirical relation describes the temperature dependence of the Soret coefficient of alkali halide solutions [15, 22, 39] and aqueous colloidal suspensions [40]

$$S_T(T) = S_T^\infty (1 - e^{\frac{T_0-T}{\tau}}), \qquad (3)$$

where $S_T^\infty$ is the asymptotic limit of the Soret coefficient at high temperatures, $T_0$ is the inversion temperature where $S_T$ switches sign, and the rate $\tau$ determines the strength of the temperature effect on $S_T$. Combining Eq. (2) and Eq. (3) provides the following relation

$$x_1(T) = \exp[-S_T^\infty (T + \tau e^{\frac{T_0-T}{\tau}}) + k], \qquad (4)$$

where $x_1(T)$ is the molar fraction of component 1 at temperature $T$, and $k$ is a fitting parameter, in addition to the previously defined $S_T^\infty$, $\tau$, $T_0$.

The thermodynamic formulation shows that the computation of the Soret coefficient ($S_T$) can be achieved by imposing a known temperature gradient and measuring the induced concentration gradient. This is used in the MD simulation approaches described below and employed for the analysis of thermal diffusion.

**Fig. 1**a shows the MD simulation cell adopted in the work. It contains ~4800 water molecules, ~450 NaCl ion pairs, and ~8900 quartz surface atoms. The system consists of NaCl aqueous solutions confined in a charged quartz slit nanochannel. Periodic boundary conditions are applied at all boundaries of the cell. Thermodynamic ensembles including the microcanonical (*NVE*), canonical (*NVT*), and isothermal-isobaric (*NPT*) ensemble are used in different stages of our modellings. The methodology for calculating the Soret coefficient of confined NaCl aqueous solutions is like the one presented in [15, 41, 42], where boundary-driven molecular dynamics simulations are used. The system first undergoes a relaxation stage in *NPT* ensemble to achieve the desired pressure by dynamically adjusting the cell's dimensions. After that, two regions of the interfacial solutions with widths ~0.15 nm are thermostatted, i.e., the temperature of the water molecules in these regions is regulated to prescribed higher and lower values relative to the rest of the system. These are the hot region and the cold region in **Fig. 1**a. The water molecules in the rest of the system, the ions, and the quartz surface atoms are not thermostatted. These particles regulate their temperature via interactions with the thermostatted water molecules by thermal conduction. A constant thermal gradient in the *x* direction can be created by directly conditioning the water molecules in the hot region at high temperatures ($T_H$), and thermostatting the ones in the cold region at low temperature ($T_C$). It has been found that a constant heat flux is established in a few hundred picoseconds. After equilibration, a steady-state condition of zero mass flux is reached, and a constant concentration gradient along the concomitant temperature gradient is established. The obtained composition and temperature profiles are then used to compute the Soret coefficient based on the theory introduced above. For comparison of different profiles, the zero planes of the quartz-water interface are specified as shown in **Fig. 1**b, where the average position of silicon atoms in the superficial silicon atoms (O-Si-O) and the ones in the silanol groups (Si-OH) has been adopted.

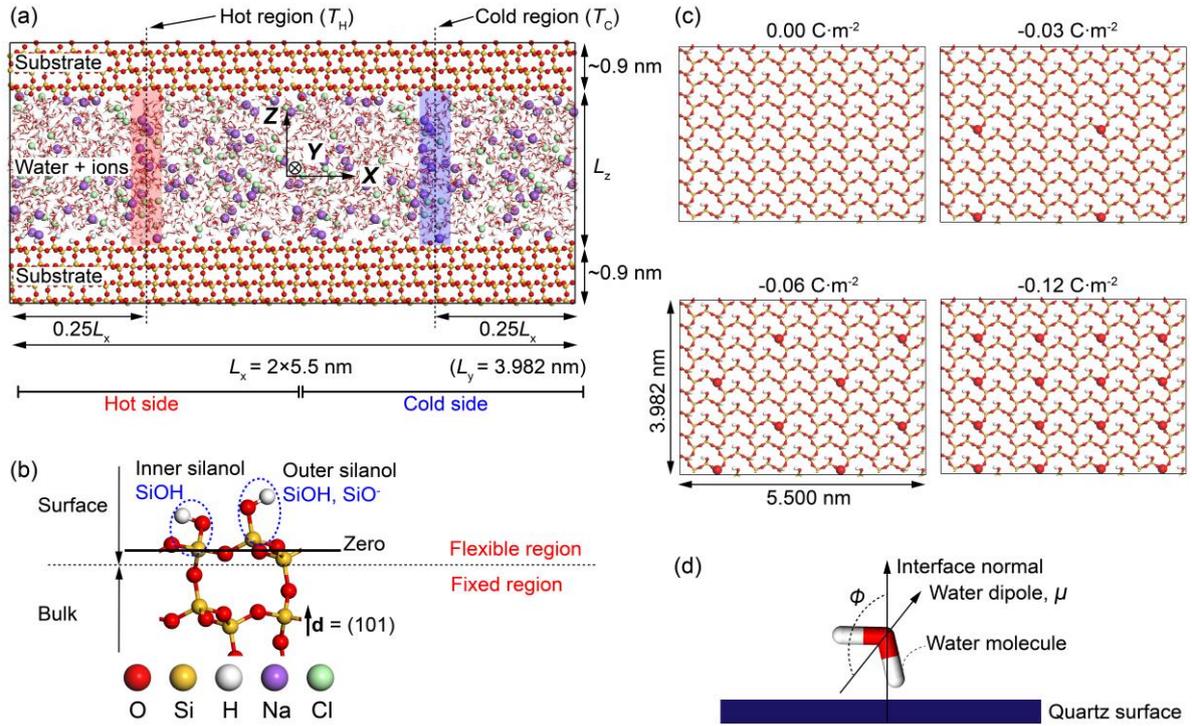

**Fig. 1.** Schematic of the MD system setup: (a) Simulation box showing dimensions of constituents and thermostatted regions; (b) Magnification focused on the quartz surface showing the inner and outer silanol groups. The solid black line denotes the zero plane, and the dashed black line separates the surface atoms (flexible region) from bulk atoms (fixed region); (c) Locations of the deprotonated outer silanol groups for the four investigated surface charge densities (Note that only half of the surface along the *x* direction is shown); and (d) Definition of $\phi$, which quantifies the spatial orientation of the water dipole with respect to the quartz surface.

It is noted that the use of solid mineral surfaces and alkali halide solutions in the MD setup of this study mimics the environments of pores in hydrothermal vents/rocks.

## 2.2 Computational details

LAMMPS [43] was adopted for all non-equilibrium molecular dynamics (NEMD) and equilibrium molecular dynamics (EMD) simulations, and the VMD [44] package was used for the trajectory visualization. The Velocity-Verlet algorithms were adopted to integrate Newton's equations of motion with a time step of 3.0 fs.

**Charged quartz substrates.** The charged quartz substrates with different surface charge densities were built by following the previous study of Kroutil et al. [45] and Quezada et al. [46]. The surface **d** = (101) of α-quartz (α-SiO$_2$) crystal is cut as the quartz-solution interface, and the under-coordinated superficial O and Si atoms are capped with H or OH groups to form outer and inner silanol groups (SiOH) (see **Fig. 1**b) as in previous studies [47, 48]. As shown in **Fig. 1**c, 0, 4, 8, and 16 outer silanol groups on each quartz surface were deprotonated by directly deleting the dangling hydrogen atoms to produce surface charge densities of approximately $0.00$, $-0.03$, $-0.06$, and $-0.12$ C·m$^{-2}$, matching along with pH values of around 2.0–4.5, 7.5, 9.5, and 11, respectively [45]. The partial charge values of the quartz surface atoms are reassigned according to the quantum mechanical calculations of Kroutil et al. [45]. The quartz interactions are described by the ClayFF [49], where the same nonbonded and bonded parameters were adopted for both uncharged and charged quartz substrates.

**NaCl aqueous solutions.** A certain amount of water molecules and sodium and chloride ions were packed into the charged quartz nanochannels to achieve an average salt concentration of 4.0 mol kg$^{-1}$ (an intermediate concentration where non-ideal effects become important [39]). This concentration enables direct comparison of the results in this work for nano-confined solutions with previous studies for bulk ones. The formed NaCl aqueous solutions were simulated utilizing the rigid SPC/E model [50] for water and the force field of Dang et al. [51-54] for the ion-ion/water interactions, whose accuracy in simulations of bulk and interfaces has been demonstrated [55]. Previous works [15, 22, 56] have shown that these force-field parameterizations accurately reproduce the experimental thermo-diffusive response of NaCl solutions, including the concentration and temperature effects on the Soret coefficient, particularly the change in sign and minimum.

The solution-quartz interactions were depicted utilizing the Coulombic and Lennard-Jones potentials, and the cross-interaction parameters between various atom types were obtained by adopting Lorentz-Berthelot rules. The cut-off distance of $r_c = 1.5$nm was adopted for the short-range interactions, and the long-range electrostatic forces were calculated utilizing the Particle-Particle-Particle-Mesh (PPPM) method [57]. RATTLE algorithm [58] was adopted to constrain the bonds involving hydrogen atoms.

**NEMD Simulation.** For NEMD simulation, first, systems were energetically minimized by adopting the steepest descent integrator. Second, a pre-equilibrium stage containing 0.1 ns *NVT* ensemble and 1 ns *NPT* ensemble was completed to reach a system pressure of $P = 600$ bar, and a system temperature of $T = (T_\mathrm{H} + T_\mathrm{C})/2$, where $T_\mathrm{H}$ and $T_\mathrm{C}$ were the thermostat temperatures in the subsequent NEMD simulations (see **Fig. 1**a). The system after this stage is referred to as the "pre-equilibrium system". Third, by switching on the two thermostats and creating a thermal gradient inside the nanochannel, the system was driven out of equilibrium, where water and ions migrate and gradually reach the steady state after several nanoseconds (9 ns in the case studies here). Finally, the production stage lasting 18 ns was run, and the configurations were collected every 100 steps (~0.3 ps) to extract the temperature, density, concentration profiles, and thermodynamic averages by separating the simulation cell into 120 statistical bins in the direction of the temperature gradient, $x$. The temperature distribution of the solution was computed based on the equipartition principle by collecting the velocities of the ions and the water molecules. Note that in the NEMD simulations, the *NVE* ensemble is applied to the system, and the temperatures of the water molecules within the two thermostatting layers were further rescaled by the local thermostats every timestep.

By applying a harmonic potential with a spring constant equal to $1000$ kJ mol$^{-1}$ nm$^{-2}$, the positions of water oxygen atoms in the hot and cold thermostatted regions were restrained in the $x$ direction, but the water molecules were allowed to move freely in the $yz$ plane. The restrained water molecules could rotate freely and exchange momentum with the unrestrained water molecules and ions. The temperatures of the hot and the cold regions were rescaled to the specific values every timestep by a canonical sampling thermostat implementing global velocity rescaling with Hamiltonian dynamics [59], and the linear momentum of the system was reset after velocity rescaling. The streaming velocity of the system was monitored, and a statistically averaged zero value was observed.

The silanol atoms, as well as the surface silicon and oxygen atoms (see **Fig. 1**b), were kept fully flexible. During the *NPT* simulations, the bulk quartz atoms were only permitted to move in the *z*-direction (fixed in the *x*- and *y*-directions) to equilibrate the system pressure and maintain the surface geometry, while in *NVT* and *NVE* simulations, these atoms were fixed in all directions (see **Fig. 1**b). These fixations were also implemented by utilizing a harmonic potential with spring constant of 1000 kJ mol$^{-1}$ nm$^{-2}$ to the fixed atoms in the corresponding directions.

**EMD Simulation.** To obtain the structural modifications of confined NaCl aqueous solutions, additional EMD simulations were performed on these systems without the thermal gradient applied. The production runs in the *NVT* ensemble lasting 20 ns were performed on the pre-equilibrium system as mentioned above to analyse the equilibration phase of the systems.

**Bulk Solutions.** To better demonstrate the effect of nanoconfinement and the surface charge on the thermo-diffusive response of the NaCl aqueous solutions, the same NEMD and EMD simulations were subsequently performed on identical NaCl aqueous solutions in bulk conditions.

## 3. Results and discussion

### 3.1 Structural modifications of nano confined NaCl aqueous solutions

At the high concentration of the solution (4.0 mol kg-1) considered, the Coulombic interactions are effectively screened [22]. Therefore, the ion-ion/water dispersive interactions become increasingly prominent. The structural properties of confined solutions were derived from the 20 ns of the trajectory at the production stage in the EMD simulations, using the charged quartz surfaces with four different surface charge densities, i.e., $0.00 \text{ C} \cdot \text{m}^{-2}$, $-0.03 \text{ C} \cdot \text{m}^{-2}$, $-0.06 \text{ C} \cdot \text{m}^{-2}$, and $-0.12 \text{ C} \cdot \text{m}^{-2}$ (see **Fig. 1**c) at the temperature of 300/350 K and the pressure of 600 bar, which is consistent with the previous experimental and numerical study [22] and facilitates the subsequent comparative analysis. At the same time, the bulk solutions under the same thermodynamic conditions were investigated for comparison.

Water molecules interact with the charged quartz surface establishing hydrogen bonds. The comparisons between the number density profiles of **Fig. 2**a–d and those of **Fig. 2**e–f show a considerable degree of overlap between solutions and the quartz surfaces, indicating penetrations of both water molecules and ions into the quartz surface. A noticeable layer-by-layer arrangement of water molecules next to the surface can be observed from the peaks and dips in **Fig. 2**a–b. This result, together with the charge density profiles of the interfacial water molecules presented in **Fig. 3**b, shows that the water molecules reorient as they meet the quartz surface, with the hydrogen atoms positioned closer to the surface than the oxygen atoms. In the middle of the nanochannel, the number density of $O_w$, $H_w$, $Na^+$, and $Cl^-$ converges to their bulk value at 300/350 K and 600 bars in about 1.2 nm from the quartz surface (see **Fig. 2**a–d).

**Fig. 2**g shows the number fraction of NaCl ions $x_{NaCl} = (N_{Na^+} + N_{Cl^-})/(N_{Na^+} + N_{Cl^-} + N_{water})$ at the charged quartz surface is significantly different from that in the interior of the nanochannel, underlining the impact of the quartz-ion interactions. This leads to the generation of an electrical double layer (EDL) in which the $Na^+$ ions absorb on the quartz surface first, resulting in a number density peak as shown in **Fig. 2**c and a number fraction peak as shown in **Fig. 2**h, while the $Cl^-$ ions absorb immediately above $Na^+$ (see **Fig. 2**d and i) with a concurrent depletion of $Na^+$ ions (see **Fig. 2**c and h). In addition, with increasing surface charge density, more cations are absorbed on the quartz surface to compensate for the negative surface charges (see **Fig. 2**c and h), while the anions are less affected by the surface charge density (see **Fig. 2**d and i). In short, the results highlight a stronger affinity of cations towards the negatively charged quartz surfaces.

The effects of nanoconfinement and surface charge on the spatial orientation of the interfacial water are quantified. The spatial orientation is characterized by $\cos(\phi)$, in which $\phi$ is the angle between the normal to the quartz surface and a vector opposite to a water dipole (see **Fig. 1**d): $\cos(\phi) = -1$ represents a water dipole pointing away from the quartz; $\cos(\phi) = 1$ represents a water dipole pointing towards the quartz; $-1 < \cos(\phi) < 1$ represents partial orientation of the water dipole. **Fig. 2**j shows the average of $\cos(\phi)$ for the interfacial water molecules at different distances from the quartz surfaces for all simulated cases. The results in **Fig. 2**j show that without electrolytes, the quartz surface induces the spatial orientation of water dipoles at the quartz-solution interface, resulting in two water layers.

The spatial orientation of water dipoles in the first layer, near the surface, is mostly pointing away from the surface ($\cos(\phi) > 0$); in the second layer, the orientation is mostly pointing toward the surface ($\cos(\phi) < 0$). Apart from the effect of nanoconfinement, **Fig. 2**j shows that the surface charge can further enhance this spatial orientation. Generally, the magnitude of $\cos(\phi)$ increases with the surface charge increasing. According to these results, the spatial orientation of interfacial water molecules, which is vital for the mobility of particles and the effective anchoring of ions, is greatly affected by the nanoconfinement and surface charge. Previous studies [42, 60] have found correlations between the thermomolecular orientation effects and the Soret effect. The results in **Fig. 2**j demonstrate that the nanoconfinement and surface charges further modify the spatial orientation of the interfacial water molecules, inducing the corresponding variations of the thermo-diffusive responses of the interfacial solutions.

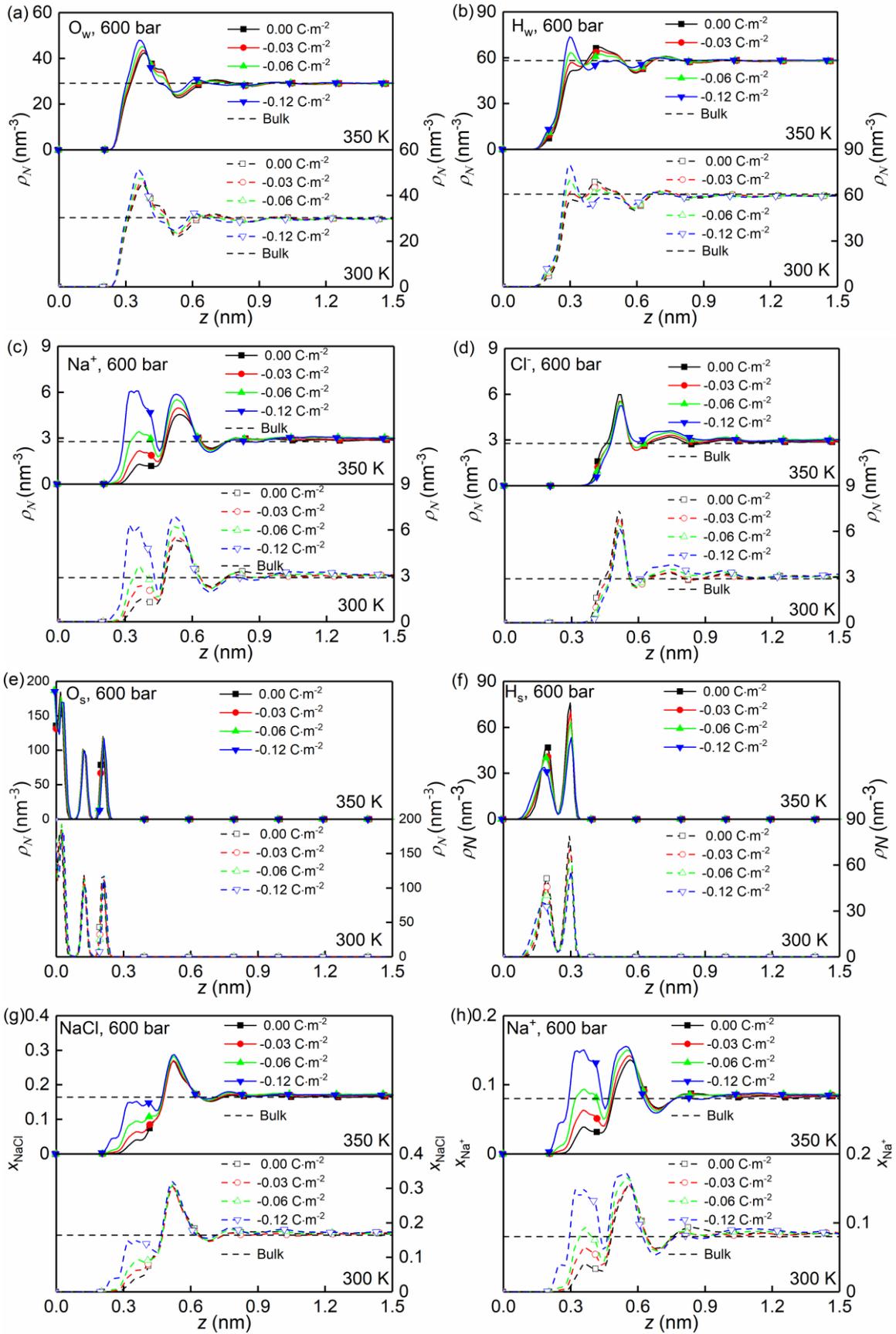

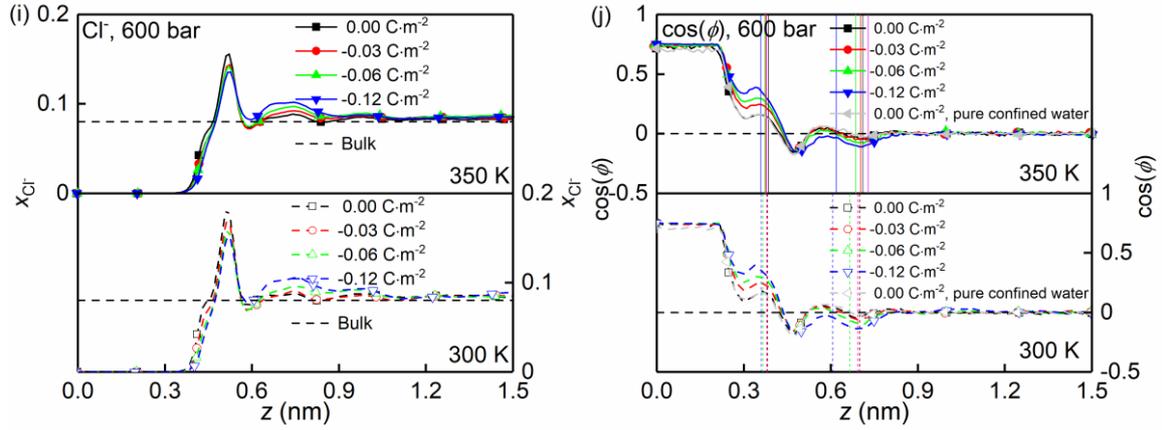

**Fig. 2.** Calculated variation of quantities in a nano-channel of width $L_z = 4.0$ nm at temperature $T = 300/350$ K and pressure $P = 600$ bar: (a–f) Number density profiles ($\rho_N$) of water oxygen atoms ($O_w$), water hydrogen atoms ($H_w$), sodium ions ($Na^+$), chloride ions ($Cl^-$), quartz surface oxygen atoms ($O_s$), and quartz surface hydrogen atoms ($H_s$) along the $z$ direction for all surface charge densities; (g–i) Number fraction of $NaCl$ ions along the $z$ direction; and (j) Water orientation on the quartz surface. Vertical solid or dashed lines indicate the positions of two water layers (peaks in the water density profiles).

The formations of electrical double layers on the charged quartz surfaces can be evaluated by the charge density distribution profiles (see **Fig. 3**). The charge separation is observed at the quartz-solution interface in **Fig. 3**a, which shows that the system reaches electroneutrality at ~1.2 nm from the quartz surface. This is consistent with the number density profiles in **Fig. 2**. The fluctuating charge density profiles (see **Fig. 3**a–d) highlight the effects of nanoconfinement and surface charge on the interfacial liquid structure.

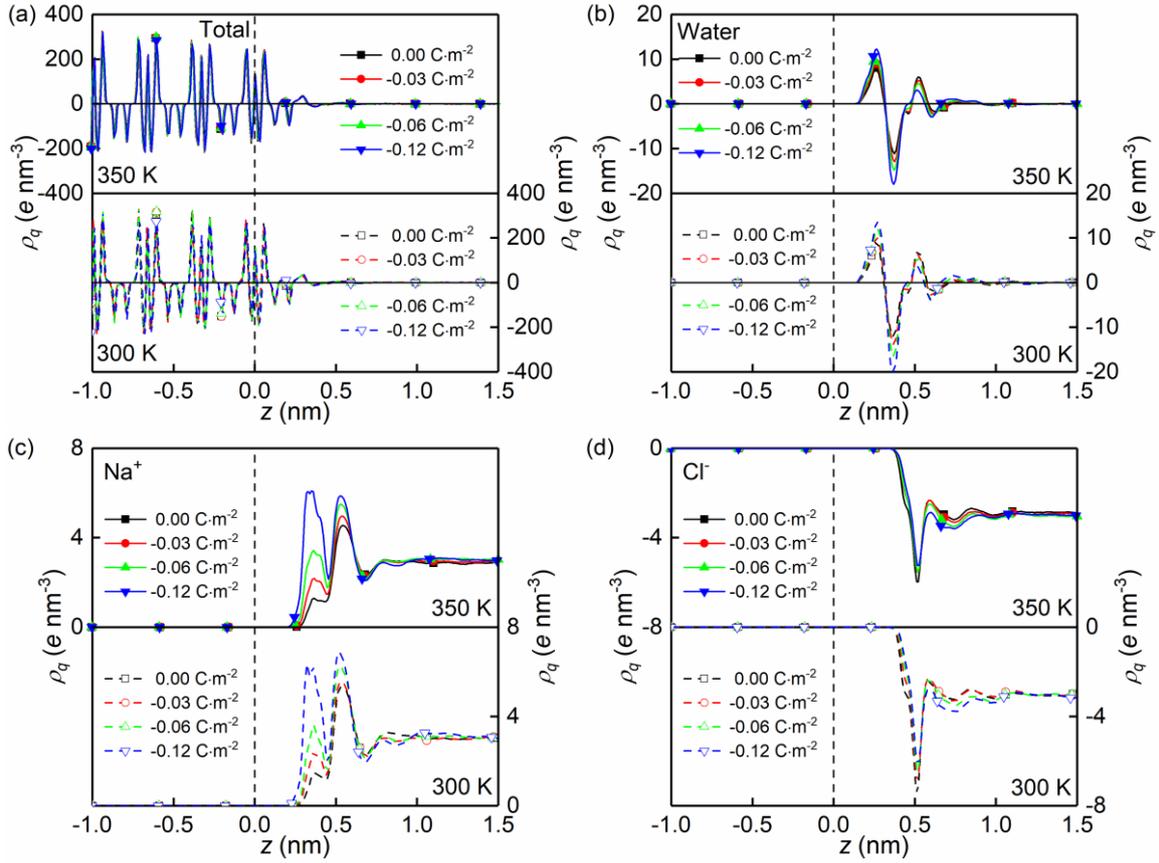

**Fig. 3.** (a–d) Charge density profiles ($\rho_q$) of the whole system, water molecules, sodium ions, and chloride ions along the $z$ direction for all surface charge densities. All the data correspond to the quartz nanochannel of width $L_z = 4.0\,\text{nm}$, temperature $T = 350\,\text{K}$, and pressure $P = 600\,\text{bar}$.

Temperature, nanoconfinement, and surface charge modify the first solvation shell of the interfacial water molecules in the NaCl solutions. These structural modifications can be quantified by the radial distribution functions (RDF), $g(r)$ (**Fig. 4**a–c and i–j), and the coordination numbers, $N_c$ (**Fig. 5**a). These show that inside the nanochannels, part of the water-water hydrogen bonds ($O_w - H_w$) are replaced by water-quartz ones ($O_w - H_s$ and $O_s - H_w$), and that the replacement is enhanced with increasing surface charge, leading to the increase of the total coordination number $N_{c,total} = N_{c,Ow-Hw} + N_{c,Os-Hw} + N_{c,Ow-Hs}$ in **Fig. 5**a. Specifically, when the system temperature ($T$) is equal to 350 K, in a system with zero surface charge, the total coordination number changes from 1.29 under the bulk conditions to 1.52 under the confined conditions, showing enhanced hydration of water molecules under confinement conditions. With increasing surface charge density, the total coordination number increases from 1.52 to 1.63 for the confined case with a surface density of $-0.12\ \text{C}\cdot\text{m}^{-2}$, highlighting that the hydration of water molecules is further enhanced by the surface charge. The hydration of water molecules can be enhanced further, albeit not strongly, by decreasing the system temperature from 350 K to 300 K.

Moreover, the (water solvation) hydration shell of the ions $Na^+$ and $Cl^-$ is modified by temperature, the nanoconfinement conditions, and the surface charges of the quartz substrates. When the system temperature ($T$) is equal to 350 K, on the one hand, the coordination number of $Na^+ - $ water (the number of water molecules around the cation $Na^+$) under the bulk conditions is 4.84 (quantified by $Na^+ - O_w$ and see **Fig. 4**e and **Fig. 5**b), while under confinement conditions and zero surface charge density, the solvation shell of $Na^+$ is slightly loosed, and the coordination number decreases to 4.77. This further decreases to 4.69 when the quartz surface charge density is $-0.12\ \text{C}\cdot\text{m}^{-2}$. On the other hand, the partly dehydrated $Na^+$ ions coordinate with the negatively charged chemical groups on the quartz surfaces, specifically, the siloxane bridges $\equiv Si - O - Si \equiv$ and the dangling oxygen $\equiv SiO^-$ (quantified by $Na^+ - O_s$ and see **Fig. 4**d and **Fig. 5**b). With surface charge density increasing from $0.00\ \text{C}\cdot\text{m}^{-2}$ to $-0.12\ \text{C}\cdot\text{m}^{-2}$, more surface oxygen atoms $O_s$ coordinate to $Na^+$ ions, leading to the increase of the total coordinate number in **Fig. 5**b. The $Cl^- - $ water coordination number also decreases under the nanoconfinement conditions, from 6.1 in bulk to 5.9 for the cases of zero surface charge (quantified by $Cl^- - H_w$ and see **Fig. 4**g and **Fig. 5**c). $Cl^-$ ions coordinate with the hydrogen atoms $H_s$ in the silanol groups of the quartz surface instead. Unlike $Na^+$ ions, the coordination number of $Cl^- - H_s$ decreases with increasing surface charge density because of the repulsive interactions between negatively charged surfaces and anions and due to less hydrogen atoms for the deprotonated silanol groups. However, no significant modifications can be observed for the total coordination number, the $Cl^- - H_w$, and $Cl^- - H_s$ coordination shells in **Fig. 5**c when surface charge density changes. In addition, an enhanced hydration shell of the ions $Na^+$ and $Cl^-$ can be observed when the system temperature is decreased from 350 K to 300 K (see **Fig. 5**b–c). In summary, a less tight solvation shell of NaCl ions can be obtained under lower temperature, nanoconfinement and surface charge conditions. This can also be seen from the different heights of the main peaks in the RDF profiles of **Fig. 4**. The position of these peaks is not significantly affected by the confinement and the surface charges. Observed also is an enhancement of ion pairing in the solution under nanoconfinement conditions or under increased temperature (see **Fig. 4**h and **Fig. 5**d). However, the variation of the surface charge appears to have little effect on ion pairing.

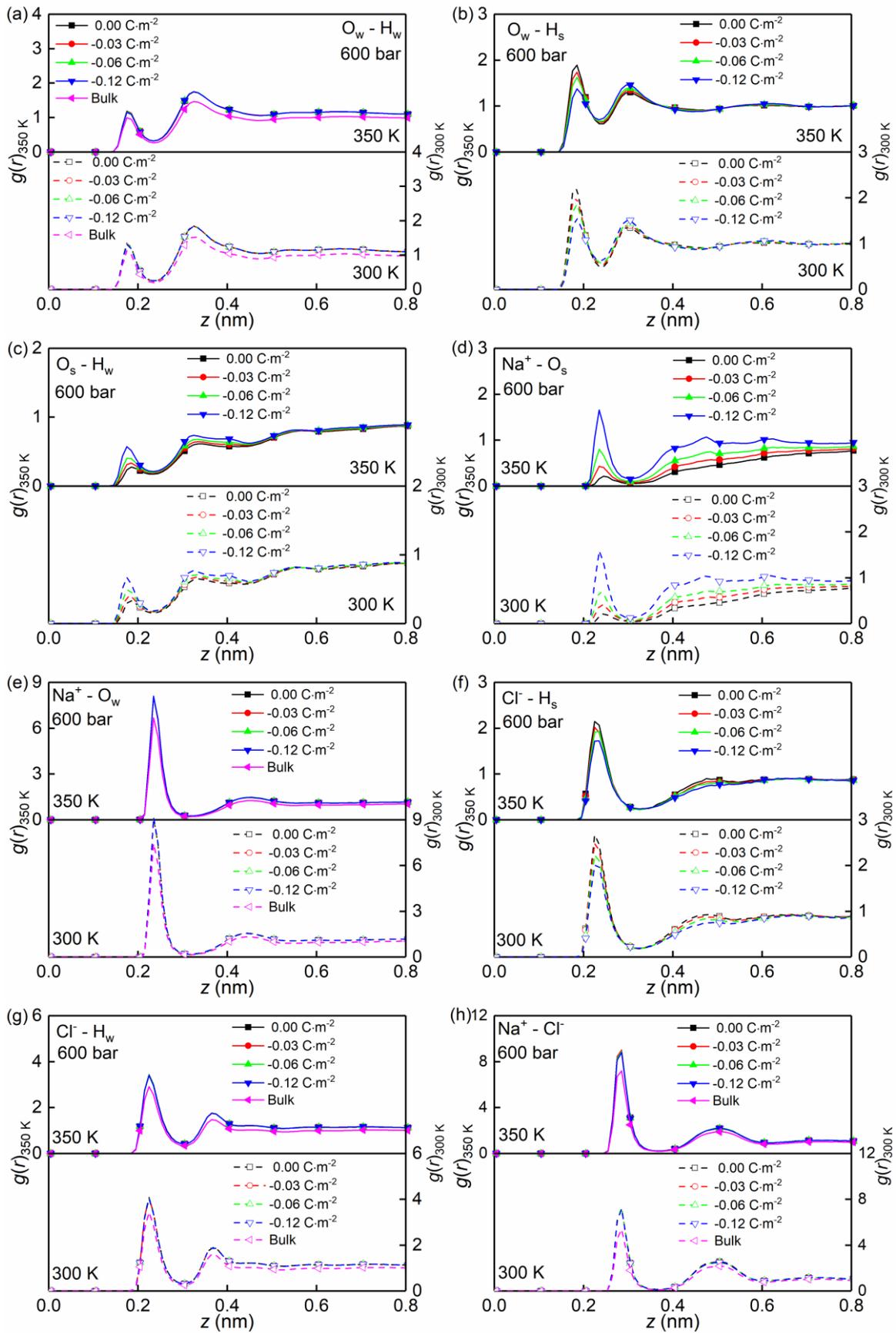
18

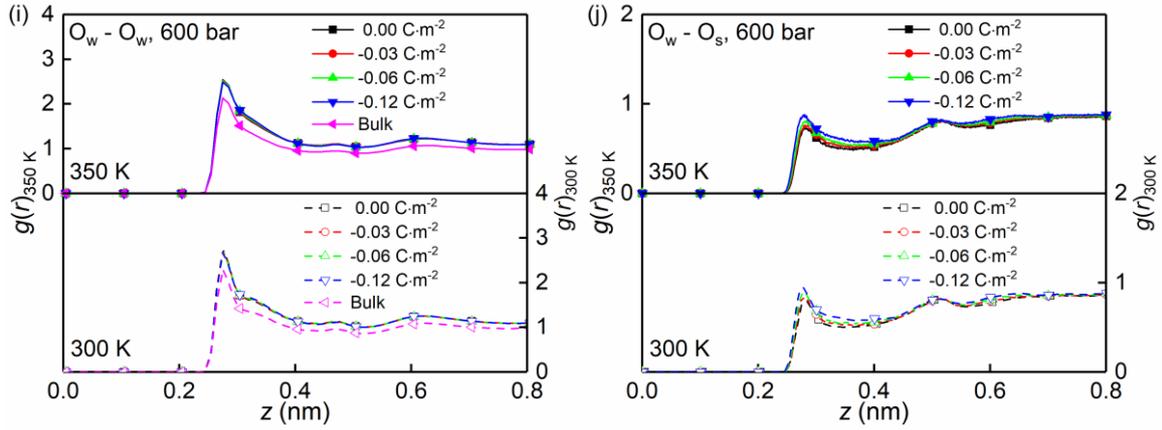

**Fig. 4.** Normalized radial distribution functions $g(r)$ for NaCl solutions confined in charged quartz nanochannels of the width $L_z = 4.0$ nm with different surface charge densities. The data correspond to 300/350 K, and 600 bar and the radial distribution functions $g(r)$ are normalized by $g(r = 1.5$nm$)$.

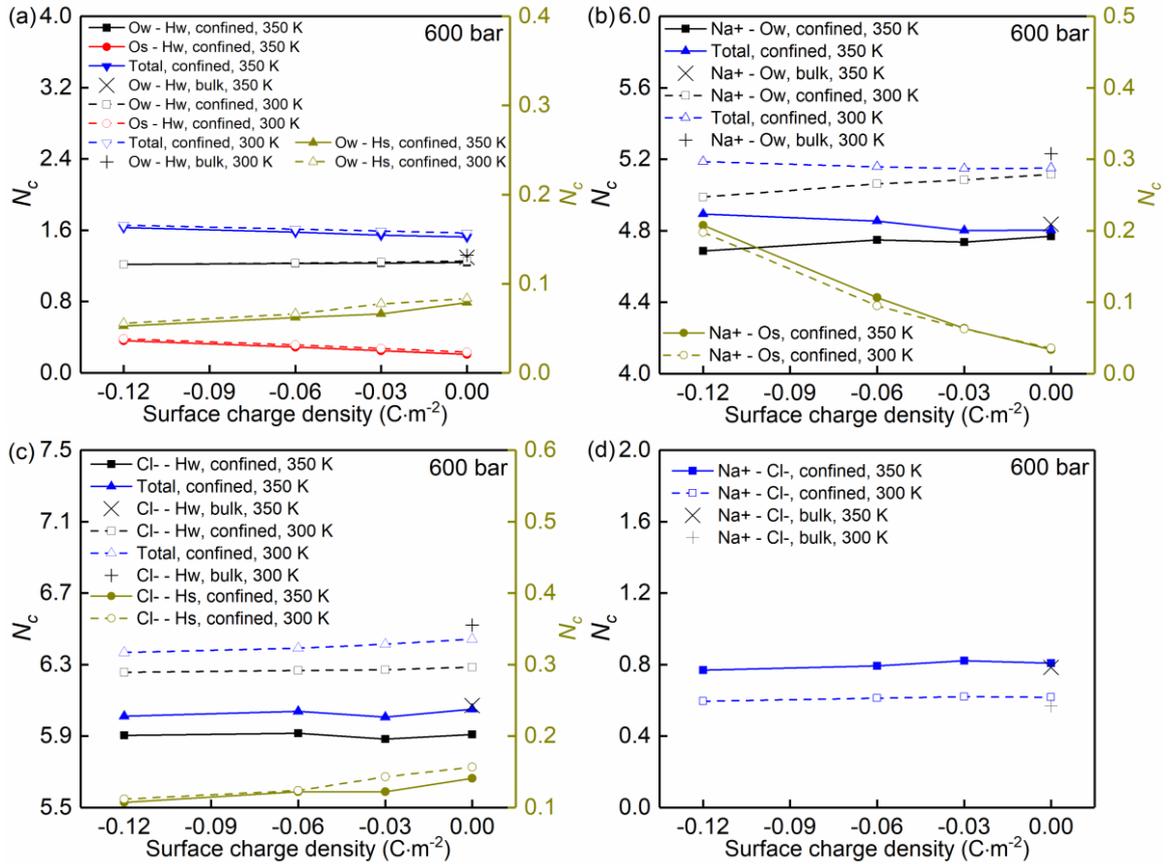

**Fig. 5.** Coordination numbers $N_c$ for NaCl solutions confined in charged quartz nanochannels of the width $L_z = 4.0$nm with different surface charge densities. The data correspond to 300/350 K, and 600 bar and the radial distribution functions $g(r)$ are normalized by $g(r = 1.5$nm$)$.

**3.2 Thermo-diffusive response of confined NaCl aqueous solutions**

Section 3.1 showed how the confinement conditions and surface charges modified the liquid structure of the solutions. This is expected to impact the thermo-diffusive response of the confined solutions. To corroborate the expectation, a temperature gradient along the $x$ direction was used in the nanochannel of width $L_z \approx 4.0$nm. The stationary state of the confined solutions was reached in ~9 ns. At this state where well-defined temperature and concentration profiles were developed. The profiles along the $x$ direction in the stationary state for different confined cases and the reference bulk condition are shown in **Fig. 6**. The concentration profiles show that the distribution of the cations is strongly affected by the surface charge density, featuring a gathering around the deprotonated silanol groups (see **Fig. 6**c), while the one of the anions is basically independent of the surface charge density (see **Fig. 6**d). This shows the screening effect of the cations on the charged quartz surfaces.

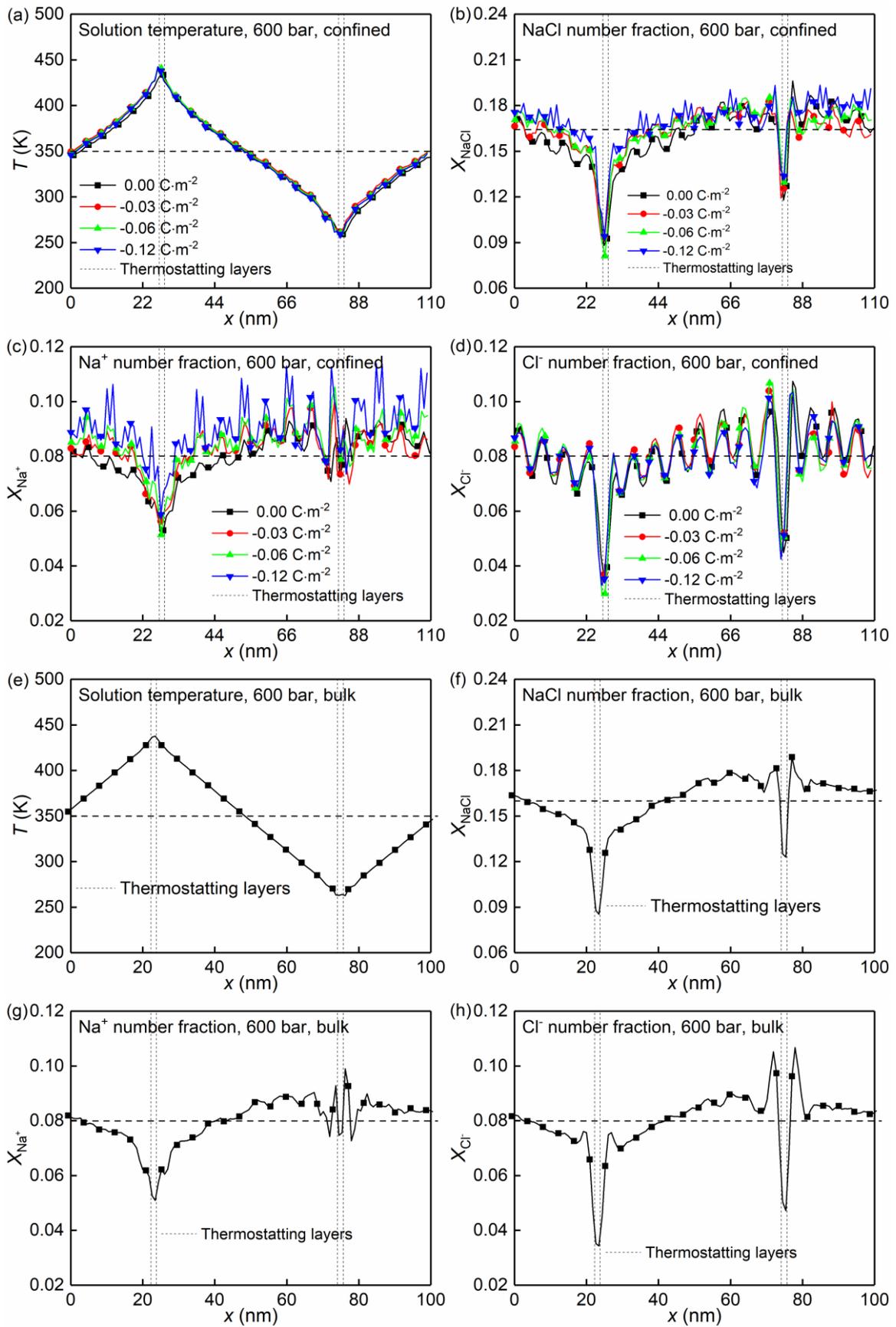
21

**Fig. 6**. Temperature and concentration profiles along the $x$ direction of the NaCl aqueous solutions inside the charged quartz nanochannels with different surface charge densities (a–d) and under the bulk conditions (e–h).

By eliminating the marginal regions where the impact of the thermostats is perceived (see the dashed thermostatted regions in **Fig. 6**), the temperature dependences of the solute molar fraction in the NEMD simulations are obtained from **Fig. 6**. These are presented in **Fig. 7**a–e by scattered symbols and fitted using Eq. (4). The fitting curves represent well the simulation results, indicated by the high $R$-square values for all cases. The obtained fitting parameters, i.e., $S_T^\infty$, $T_0$, and $\tau$ are further used in Eq. (3) to calculate the Soret coefficients of the confined NaCl aqueous solutions at different temperatures. These are presented in **Fig. 7**f. Romer et al. [22] have reported experimental and simulated Soret coefficients of bulk NaCl aqueous solutions at similar thermodynamic states and concentrations, which are also given in **Fig. 7**f. The Soret coefficients calculated here agree with the simulated values in ref [22], which verifies the accuracy of the results in this work. The calculated Soret coefficients show the same temperature dependence as the ones obtained by experiments but with lower magnitudes. Currently, the existing MD simulation models for NaCl aqueous solutions can only reproduce Soret coefficients of the right order of magnitude and predict the correct dependence of the Soret coefficient with temperature and concentration, possibly due to the contradiction between the complexity of this non-isothermal transport phenomenon and the comparative simplicity of the interatomic interaction model. This discrepancy between MD and experimental results within $\sim 10^{-3}$ K has been thus deemed as acceptable in previous studies [15, 22]. Indeed, the Soret coefficients measured by different experimental approaches under the same thermodynamic conditions also disagree with each other (e.g., the infrared thermal diffusion forced Rayleigh scattering experiment [22] and thermo-gravitational column experiments [61]). Compared with the results in the bulk solutions, the measured Soret coefficient of the confined solutions in charged nanochannel is generally lower, indicating that the solution becomes more thermophilic. It is found that after the equilibrium stage, the steady states of the thermal gradient and concentration gradient have been reached, with respective characteristic times of 1 and 9 ns. The position of the maximum in the molar fraction profile (see **Fig. 7**a–e) indicates the sign-reversal temperature, $T_0$, where the Soret coefficient is zero. It is found that this sign-reversal temperature increases with the surface charge density. All cases studied here show a temperature inversion effect. The sign of the Soret coefficient changes at the temperature of $T_0$. At $T_0$ the thermo-diffusive response of the solution shifts from thermophilic (the salt accumulates in the cold region) at low temperature to thermophobic (the salt accumulates in the hot region) at high temperature. In addition, it is also found in **Fig. 7**f that the solution tends to be more thermophilic with increasing negative surface charges. The previous work by Di Lecce et al. [41] has shown that by varying the nanopore size, the thermal diffusion of alkali halide solutions in silica nanopores can be effectively regulated. The results of the present work show, for the first time, that changing the surface charge density can also achieve the same effect.

The role of hydrogen bond structure in thermal diffusion has been previously discussed by Niether and Wiegand [62]. Their conclusions show that water-water hydrogen bonds are easier to form at low temperatures, and the solution is more thermophilic, while the water-water hydrogen bonds are destroyed at high temperatures, and the solution is more thermophobic. These conclusions agree well with the results for bulk NaCl solutions (see **Fig. 7**f and h), namely, under high-temperature conditions, the quantity of hydrogen bonds is lower in the bulk solution, which therefore is more thermophobic. For solutions under nano-confinement and surface charge, the total quantity of hydrogen bonds is higher than that in bulk, which agrees well with the observation in **Fig. 7**f that under nanoconfinement and surface charge, the solution exhibits a more intense thermophilic behaviour, with the ions tend to accumulate in hot areas.

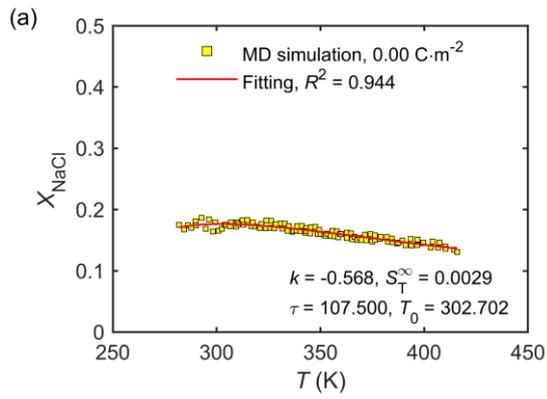
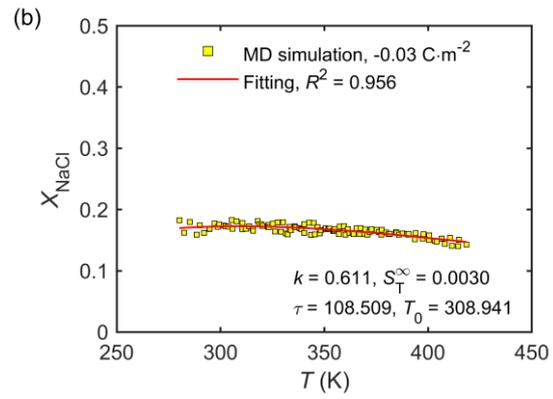
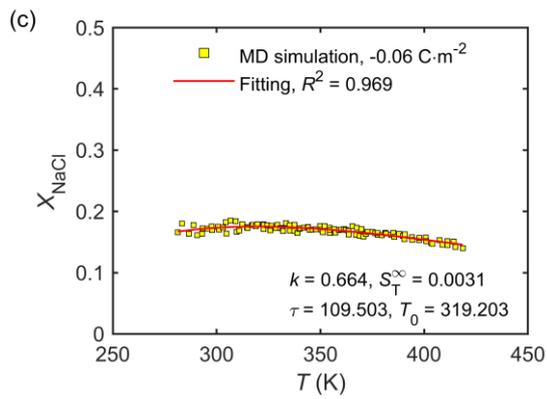
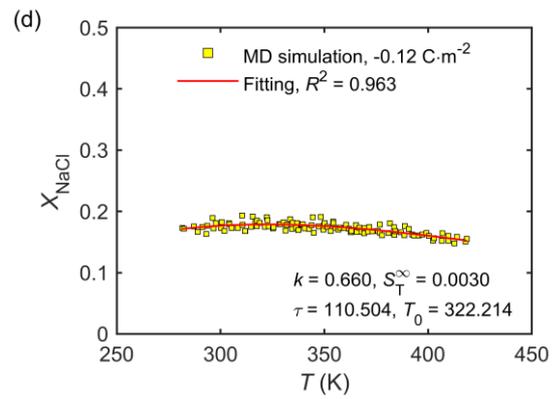
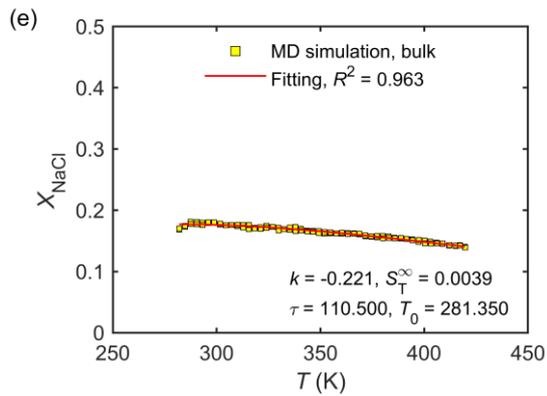
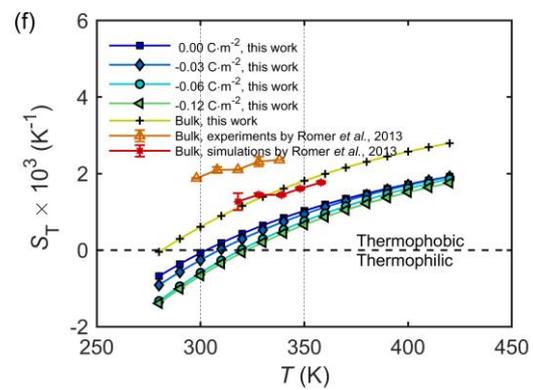
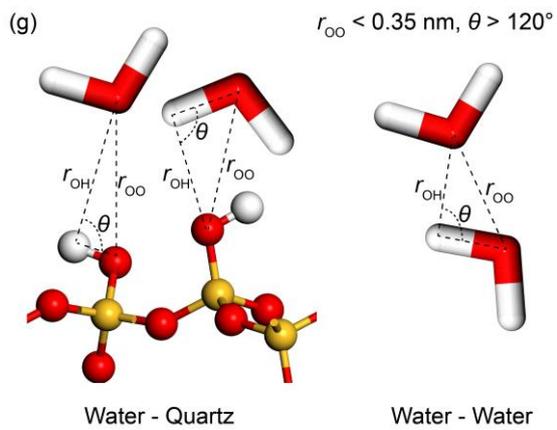
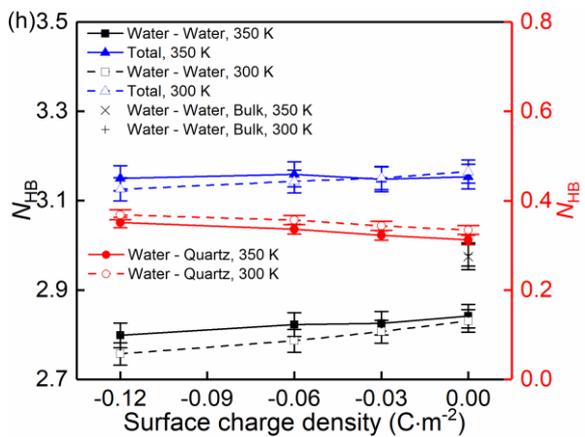

**Fig. 7.** (a–e) Temperature dependence of the molar fraction of NaCl ions at the stationary state in the bulk and confined in the charge quartz nanochannels, where full lines are using equation (4). (f) Temperature dependence of the Soret coefficient for the NaCl solutions in the bulk and confined in the charge quartz nanochannels. Our MD simulations used a thermal gradient of $\nabla T = 34$ K/nm and the average temperature, pressure, and salt molality were set to 350 K, 600 bar, and 4.0 mol/kg, respectively. (g) The adopted criterion to define the hydrogen bonds between water molecules and silanol groups, as well as water molecules and water molecules. (g) Average hydrogen bond number per water molecule in NaCl solutions confined by charged quartz nanochannels and in bulk.

It has been shown previously that the hydration structure and entropy of the ions have a considerable influence on the magnitude of the Soret coefficient [39]. Their conclusion is that a less tight water solvation shell of NaCl ions makes the solution more thermophilic and results in a more negative Soret coefficient. This conclusion is in full agreement with the results in **Fig. 5** and **Fig. 7**f and the associated discussions.

**Fig. 8** shows the molar fraction distributions of the NaCl along the $z$ direction in the hot and cold side (see **Fig. 1**a) for four kinds of charged quartz slit pores. It is found that the thermo-diffusive responses of the NaCl aqueous solutions are different in the bulk liquid and in the boundary layers, characterized by the differences between the hot and the cold sides (solid grey lines). The difference is constant in the bulk liquid and fluctuating in the boundary layers. Secondly, the positive value of the difference indicated that the NaCl is driven from the hot toward the cold side (thermophobic, $S_T > 0$) while the negative one means that the NaCl is migrating from the cold toward the hot side (thermophilic, $S_T < 0$). Therefore, it is found that with the surface charge increasing, the thermo-diffusive behaviour of some regions in the boundary layers is thermophilic instead of thermophobic, as shown in **Fig. 8**d.

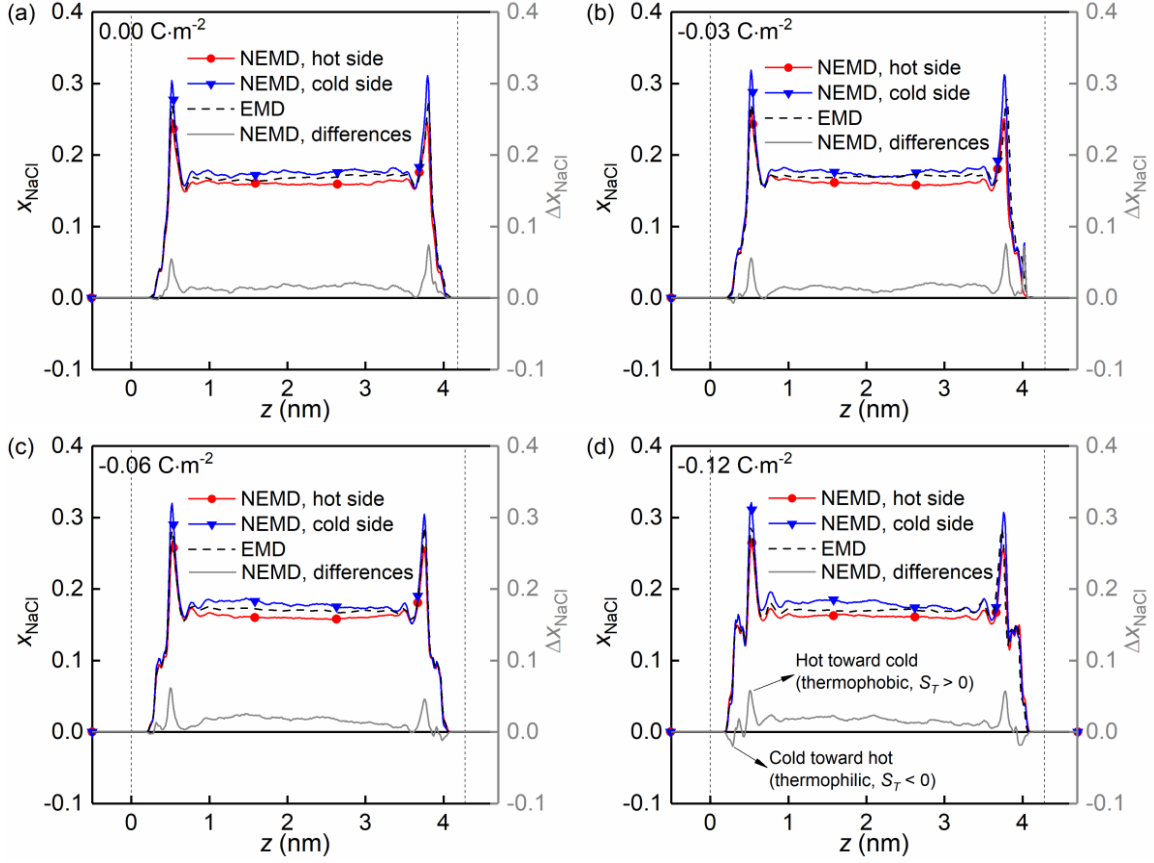

**Fig. 8.** (a–d) Evolution of the molar fraction distribution of the NaCl along the $z$ direction in the hot and cold side for four kinds of surface charges.

Using the atomic velocities from EMD simulations, the one-dimensional effective mutual diffusion (interdiffusion) coefficient in the $x$ direction, i.e., $D_{12}$ in Eq. (2), was quantified by the time integral of the autocorrelation function of the concentration current [63, 64],

$$D_{12} = \frac{1}{(N_1+N_2)x_1 x_2} \int_0^\infty \langle J_{12,x}(0) J_{12,x}(t) \rangle \, \mathrm{d}t, \qquad (5)$$

where $N_1$ and $N_2$ are the particle numbers of component 1 (NaCl) and component 2 (water), $x_1 = N_1/(N_1 + N_2)$ and $x_2 = N_2/(N_1 + N_2)$ are their molar fractions,

$$J_{12,x}(t) = x_2 \sum_{i=1}^{N_1} v_{i,x}(t) - x_1 \sum_{j=1}^{N_2} v_{j,x}(t), \tag{6}$$

is the relative velocity of concentration current in the $x$ direction, and $v_{i,x}(t)$ and $v_{j,x}(t)$ are the particle velocities. The calculated results are presented in **Fig. 9**a. They show that both nanoconfinement conditions and surface charges can lower the mutual diffusivity. Previous studies [26-29, 65, 66] have attributed this phenomenon to structural modifications, e.g., the extension of the size of electrical double layers (EDL) ion distribution. In addition, the calculated $D_{12}$ enables evaluation of the thermal diffusivity of NaCl ions using Eq. (2), i.e., $D_T T = S_T D_{12} T$, where, $S_T$ is presented in **Fig. 7**f. The calculated $D_T T$ is presented in **Fig. 9**b. It shows that the thermal diffusivity of NaCl ions is slightly lower than that of concentration-driven ionic diffusivity, i.e., $10^{-10} \sim 10^{-9} m^2/s$. This suggests that thermal diffusion should not be neglected when considering coupled nanoscale transport under temperature gradients. Moreover, it is found that, like ionic diffusivity, the thermal diffusivity of NaCl ions is lowered by nano-confinements and surface charges. In summary, based on the existing studies on concentration-driven diffusive transport in nanofluidic systems, this work shows that thermally-induced diffusive transport is affected by nano-confinement and surface charges, which facilitate the structural modifications of EDL.

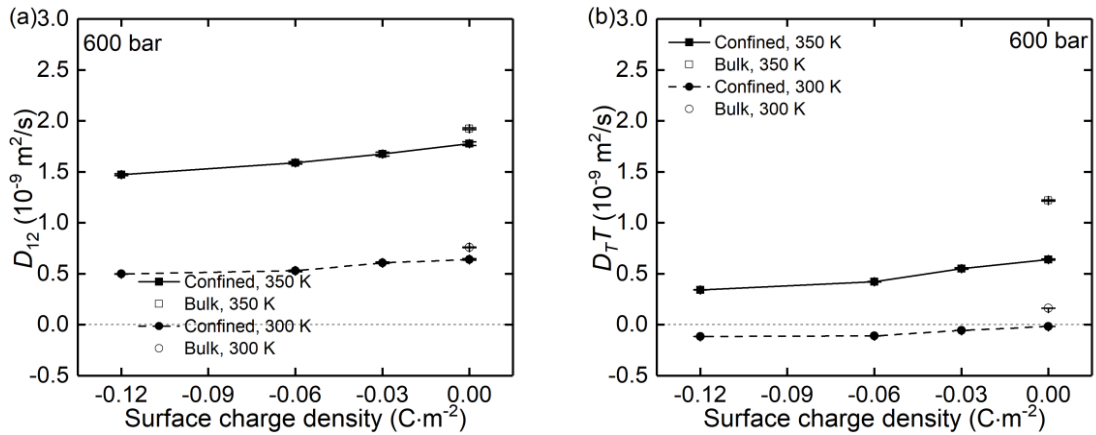

**Fig. 9.** The (a) mutual diffusion coefficients ($D_{12}$) and (b) thermal diffusivity ($D_T T$) of NaCl ions under different surface charge conditions and at temperatures of 300 K and 350 K.

## 4. Conclusions

Molecular dynamics (MD) simulations were performed to investigate the thermal diffusion of NaCl solutions in quartz nanochannels with surface charge densities ranging from $0.00 \text{ C} \cdot \text{m}^{-2}$ to $-0.12 \text{ C} \cdot \text{m}^{-2}$. The results obtained with the non-equilibrium molecular dynamics (NEMD) analyses showed that the nano-confinement and the surface charges considerably reduced the thermo-phobic response of 4.0 mol/kg NaCl solutions. It was found that by increasing the surface charge density, the ionic solution became thermophilic, reflected in a negative Soret coefficient ($S_T < 0$), i.e., the salt moved from cold to hot regions. The equilibrium molecular dynamics (EMD) simulations revealed that the structural modifications induced by the nano-confinement and the surface charge were strongly related to the observed thermo-diffusive response of NaCl solution in the NEMD simulations, and the thermal diffusion behaviour was different in the bulk liquid and the boundary layers.

The study revealed for the first time that the thermal diffusion of alkali solutions in nanopores can be controlled not only by varying the nanopore size, but also by changing the surface charge density. This outcome is important for the design of many devices/facilities, such as nanofluidic separation devices that utilise thermal fields to separate and enrich salts, nano-porous membranes for water desalination and low-grade waste heat recovery, engineering barriers for hazardous contaminants and high-level nuclear waste etc. The modelling methods presented in this paper, together with the insights derived from our results, show a path for further investigation of coupled transport phenomena in nanoscale structures.

**Author contributions**

W. Q. C., M. S., and A. J. conceived the idea. W. Q. C. performed the numerical simulations. All authors discussed the results and wrote and reviewed the manuscript. M. S. and A. J. supervised the entire project.

**Declaration of Competing Interest**

The authors declare that they have no known competing financial interests or personal relationships that could have appeared to influence the work reported in this paper.

**Acknowledgement**

Chen acknowledges the President Doctoral Scholarship Award (PDS Award 2019) from The University of Manchester, UK. Jivkov acknowledges gratefully the financial support from the Engineering and Physical Sciences Research Council (EPSRC), UK, via Grant EP/N026136/1. The authors acknowledge the assistance provided by the Research IT team for the use of the Computational Shared Facility at The University of Manchester.